\documentclass[sigconf]{acmart}

\usepackage{listings}
\usepackage{parcolumns}
\usepackage{xcolor}
\usepackage{tikz}
\usepackage{algorithm}
\usepackage{algorithmic}

\usetikzlibrary{tikzmark,positioning}
\AtBeginDocument{%
  }

\setcopyright{acmcopyright}
\copyrightyear{2018}
\acmYear{2018}
\acmDOI{XXXXXXX.XXXXXXX}

\acmConference[Conference acronym 'XX]{Make sure to enter the correct
  conference title from your rights confirmation emai}{June 03--05,
  2018}{Woodstock, NY}
\acmPrice{15.00}
\acmISBN{978-1-4503-XXXX-X/18/06}

\begin{document}

\title{Smart Contract Upgradeability on the Ethereum Blockchain Platform: An Exploratory Study}

\author{Ilham Qasse}
\affiliation{%
  \institution{Reykjavik University}
  \city{Reykjavik}
  \country{Iceland}}
  \email{ilham20@ru.is}
\author{Mohammad Hamdaqa}
\affiliation{%
  \institution{Polytechnique Montreal}
  \city{Montreal}
  \country{Canada}}
\email{mhamdaqa@polymtl.ca}
\affiliation{%
  \institution{Reykjavik University}
  \city{Reykjavik}
  \country{Iceland}}
\email{mhamdaqa@ru.is}
\author{Björn Þór Jónsson}
\affiliation{%
  \institution{Reykjavik University}
  \city{Reykjavik}
  \country{Iceland}}
  \email{bjorn@ru.is}

\renewcommand{\shortauthors}{Trovato et al.}

\begin{abstract}
\textit{\textbf{Context:}}Smart contracts are computerized self-executing contracts that contain clauses, which are enforced once certain conditions are met. Smart contracts are immutable by design and cannot be modified once deployed, which ensures trustlessness. Despite smart contracts' immutability benefits, upgrading contract code is still necessary for bug fixes and potential feature improvements.
In the past few years, the smart contract community introduced several practices for upgrading smart contracts. Upgradeable contracts are smart contracts that exhibit these practices and are designed with upgradeability in mind. During the upgrade process, a new smart contract version is deployed with the desired modification, and subsequent user requests will be forwarded to the latest version (upgraded contract). 
Nevertheless, little is known about the characteristics of the upgrading practices, how developers apply them, and how upgrading impacts contract usage. 

\textit{\textbf{Objectives:}}  This paper aims to characterize smart contract upgrading patterns and analyze their prevalence based on the deployed contracts that exhibit these patterns. Furthermore, we intend to investigate the reasons why developers upgrade contracts (e.g., introduce features, fix vulnerabilities) and how upgrades affect the adoption and life span of a contract in practice.

\textit{\textbf{Method:}} We collect deployed smart contracts metadata and source codes to identify contracts that exhibit certain upgrade patterns (upgradeable contracts) based on a set of policies. Then we trace smart contract versions for each upgradable contract and identify the changes in contract versions using similarity and vulnerabilities detection tools. Finally, we plan to analyze the impact of upgrading on contract usage based on the number of transactions received and the lifetime of the contract version.

\end{abstract}



\maketitle

\section{Introduction}

A smart contract is a computer code deployed onto the blockchain to enforce, monitor, and execute agreements when conditions are met~\cite{wang2019blockchain}. Smart contracts are immutable by design, ensuring trustlessness as they effectively serve as an unbreakable contract between participants. 

Software quality, however, depends upon the ability to update and patch source code to accommodate identified vulnerabilities and bugs after release. 
Studies show that several deployed smart contracts have been exposed to attacks because of vulnerabilities and bugs in the contract code~\cite{he2020smart,qian2022smart, samreen2021survey}.
In several cases, millions of dollars have been stolen due to vulnerabilities in smart contract code~\cite{he2020smart,qian2022smart, samreen2021survey}. Despite smart contracts' immutability benefits, upgrading contract code is still necessary for bug fixes and potential feature improvements ~\cite{antonino2022specification,salehi2022not}.

In contrast to traditional software, where a software upgrade can introduce changes to any part of the software, in smart contracts, a contract upgrade refers to the process of arbitrarily modifying the contract code while maintaining contract state~\cite{antonino2022specification,salehi2022not}.

In the past few years, the smart contract community introduced several practices for upgrading smart contracts.\footnote{\url{https://eips.ethereum.org/EIPS/eip-1822}}\footnote{\url{https://eips.ethereum.org/EIPS/eip-2535}}\footnote{\url{https://eips.ethereum.org/EIPS/eip-3448}} 
Since smart contracts are immutable by design, not all deployed smart contracts are upgradeable smart contracts. An upgradeable smart contract is a contract that incorporates upgrading approaches proposed by the community and is designed specifically for possible upgrading. 
However, the smart contract upgrading approaches are relatively new, and there is a limited understanding of how they are applied. Furthermore, despite the increasing interest in smart contracts and mining their repositories, analyzing smart contract versions and the impact of upgrading smart contracts (for example, on activity level) is an open research topic. 

This empirical study aims to identify upgradeable smart contracts that exhibit certain upgrading approaches and analyze the prevalence of these upgradeable contracts, including the commonly applied upgrading approaches. We will focus on upgrading patterns introduced by the Ethereum community. \footnote{\url{https://ethereum.org/en/developers/docs/smart-contracts/upgrading/}} 
Moreover, we will study the upgrade patterns through the lens of evolution lineage (different versions of deployed smart contracts) to answer the following Research Questions (RQs): 

\begin{itemize}
\item \textbf{RQ1: How prevalent are upgrading patterns in smart contracts?}

\textbf{Motivation:} Although the smart contract community proposed many approaches to upgrade smart contracts, little is known about the adoption of these approaches by developers, as well as which approach is commonly used in practice. Answering this research question will provide insights into the prevalence of the various upgrade approaches. Moreover, it might help developers and researchers identify trends and best practices for designing contracts with upgradeability in mind. We will design policies to identify upgradeable contracts and the approach used to upgrade them. These policies include a combination of regular expressions specific to the upgrading pattern as well as the code structure of the pattern. 


\item \textbf{RQ2: How likely is an upgradeable contract to be upgraded?}


\textbf{Motivation:} While utilizing upgrade patterns in designing a contract provide the possibility to upgrade the contract at a later stage of its life cycle. However, it is unclear what percentage of contracts with this capability really end up being upgraded.  
To answer this question, we will trace deployed smart contract versions for each upgradeable contract. We will achieve this by obtaining the historical records for these contracts and identifying whether an upgrade occurs during the contract’s lifetime. The results of this question include a list of smart contract versions (if upgraded), and their addresses. These results can be used to understand better whether upgrading contracts is an adopted practice in the smart contract community, and get better insights on the importance of designing contracts with upgradability in mind. Moreover, This may also help to uncover other patterns of upgrading smart contracts.

\item \textbf{RQ3: How are upgrade patterns applied in practice(e.g., improve security, introduce features, update contract policies)?}

\textbf{Motivation:} There is a lack of studies that analyze how upgrading approaches are applied in practice and the rationale behind upgrading a contract. 
To answer this question, we will analyze the evolution of smart contracts and identify the most common changes between smart contracts' versions. The results might help developers and researchers understand smart contracts' overall evolution and maintenance practices and make informed decisions about how to maintain and update contracts. 

\item \textbf{RQ4: How does smart contract upgrading impact the activity level of the contract?}

\textbf{Motivation:} Smart contract upgradeability may lead to a trust concern for smart contract users. This is because many users trust contracts because of their immutability. However, smart contract upgradeability breaks the immutability of the contracts. Thus, this can reduce the number of users of the smart contract using it.
Moreover, when a smart contract is upgraded, its internal logic and behavior may change, affecting how users interact with the contract. Users may need to adapt to changes in the contract's interface, and the gas consumption of transactions may also change. Additionally, if a smart contract is upgraded frequently or without proper communication to users, it may lead to a loss of trust and decreased activity.

Analyzing the impact of upgrading a smart contract on its activity level enables developers and stakeholders to understand how upgrading a contract will affect its usage and overall adoption from its users. This can be useful for planning future developments and upgrades for the contract.
Hence, in this research question, we analyze the impact of upgrading a smart contract on its activity level, considering each version's different lifetimes.
\end{itemize}

The reminder of the paper is organized as follows. Section~\ref{Bac} provides an overview of smart contracts and their upgradability. Section~\ref{SM} discusses the proposed research methodology to answer the research questions. We present the execution plan and premilinary results in Section~\ref{EPPR}. Section~\ref{RW} presents the related work. Finally, Section~\ref{TH} discusses research challenges and limitations.

\section{Background}
\label{Bac}
This section introduces concepts related to this study's research questions, including an overview of smart contracts and their upgradeability.
\subsection{Smart Contract}
Smart contracts are computer programs that auto-execute agreements when certain conditions are satisfied. Smart contracts are mostly deployed on a blockchain network, enabling the execution of the contract to be secure and transparent. 
Many blockchain platforms support smart contracts, such as Hyperledger Fabric, Neo, Corda, and Ethereum. This paper focuses on the smart contracts deployed on Ethereum due to the platform's open-source architecture and community. Ethereum is an open-source platform, meaning the underlying code and protocols are publicly available. Moreover, it has a large active community that supports researchers and developers with smart contracts. 

Ethereum runs its deployed smart contracts on the Ethereum Virtual Machine (EVM), a low-level stack machine that executes the compiled bytecode of the smart contract.
Each operation in Ethereum requires a certain amount of computational effort, measured by gas. Gas is required for every operation in Ethereum, whether the operation is a transaction or the execution of a smart contract instruction. Some instructions are gas-intensive, such as instructions that utilize replicated storage. Gas metering prevents the sender from wasting computational power in executing unnecessary computation-intensive transactions. Moreover, it limits the number of instructions a transaction can execute, preventing non-terminating executions and DoS attacks.

Smart contracts in Ethereum are developed in many languages, such as Solidity, Vyper, and Bamboo. In this study, we target Solidity smart contracts, the most popular smart contract programming language~\cite{clack2016smart}. 
The Solidity language is an object-oriented language deployed on the EVM. In addition to Ethereum, several blockchain platforms support Solidity, including Quorum, Hyperledger Burrow, and Hyperledger Besu. Solidity has syntax similar to C and JavaScript, but it includes several unique concepts specific to smart contracts, including: visibility of function modifiers; internal, external, view; emitted events; and smart contract-specific operations such as self-destruct and revert.

We focus on solidity contracts as there are more than 44 million solidity contracts deployed on the Ethereum network. Additionally, Solidity supports libraries, which are useful for implementing reusable code in smart contracts. This might be beneficial in upgradeable smart contracts, as libraries can be modified and updated without requiring the redeployment of the entire contract.

\subsection{Smart Contract Upgradeability}
\label{SMU}
In smart contracts, upgradeability refers to the process of modifying smart contract code after deployment while maintaining the contract data and state ~\cite{antonino2022specification,salehi2022not}. 
Upgrading smart contracts has two benefits: (i)~it provides a mechanism to improve contract security by fixing security issues and bugs discovered post-deployment, and (ii)~it enables developers to add new features and functionality over time~\cite{antonino2022specification,salehi2022not}.

In the context of smart contracts, upgradeability and mutability (ability to change) are different. Since smart contracts are immutable by design, their code cannot be changed once deployed. However, the smart contract community proposed several mechanisms to upgrade the contract, such as deploying a new smart contract and directing user requests to the new one instead of the previous one. 

An example is the proxy pattern, where users interact with a proxy contract instead of the business logic contract. The proxy contract stores data and forward user requests to the targeted smart contract version. Whenever there is an upgrade, the admin deploys new smart contracts and specifies its address in the proxy contract as the target version.
Different proxy patterns exist, such as Diamond Proxy and Universal Upgradeable Proxy Standard (UUPS). These variants follow the same concept but differ in storage and code structure. 

Another example is the data separation pattern,  where there are two contracts: a storage contract and a business logic contract. This pattern follows a similar structure as the proxy pattern, but here the logic contract (different versions) calls the storage contract. In this pattern,  users interact with the logic contract while the storage contract provides access to data only to authorized contract versions.
Other upgrading approaches exist, such as the strategy pattern and data migration approach. In this study, we will only focus on the upgrading approaches introduced by the Ethereum community. 

\section{Study Method}
\label{SM}
\begin{figure}[t]
    \centering
    \includegraphics[scale=0.52]{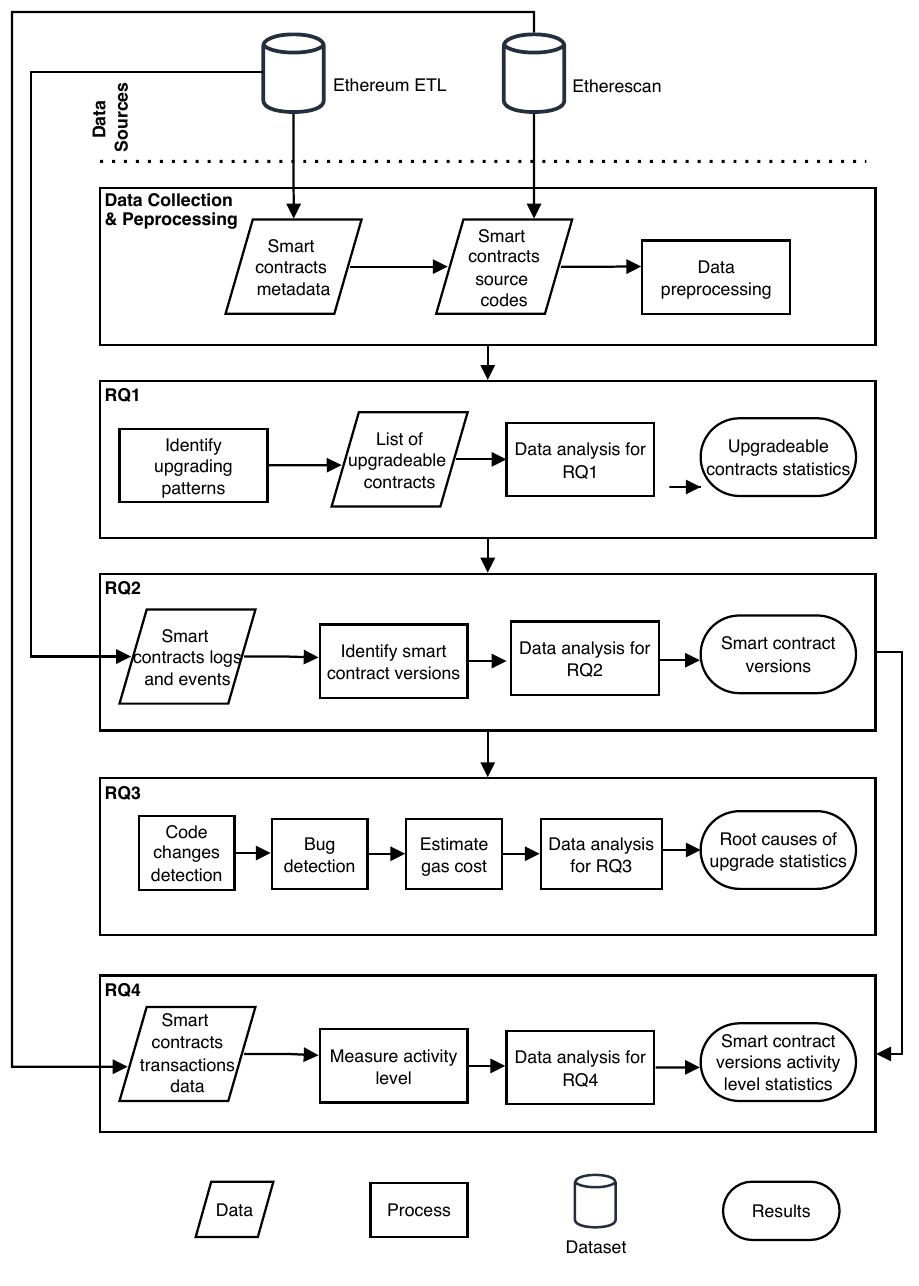}
    \caption{Overview of the study method}
    \label{fig:RO2}
\end{figure}

This study aims to empirically mine the upgradeable smart contract patterns from the deployed smart contracts to provide evidence of their prevalence and usage scenarios. Figure \ref{fig:RO2} illustrates an overview of our study methods, including data collection, data preprocessing,  identifying upgradeable smart contracts, and analyzing smart contracts version evolution and their impact on the contract's activity level.
\subsection{Data Sources}
\label{DC}
In order to conduct our study, we will integrate data from two sources, which are Ethereum ETL,\footnote{\url{https://ethereum-etl.readthedocs.io/en/latest/}} and Etherscan.\footnote{\url{https://etherscan.io}} 

Ethereum ETL is a public Ethereum data explorer that enables users to explore blockchain data such as blocks and transactions. All the data is published as a Google BigQuery dataset.\footnote{\url{https://bigquery.cloud.google.com/dataset/bigquery-public-data:crypto_ethereum}}The Ethereum ETL dataset includes details about blocks, transactions, and smart contracts. Ethereum ETL configures nodes to synchronize data with the Ethereum blockchain, where these data are updated regularly. We will extract from the dataset smart contract metadata such as smart contract address, bytecode, and the contract creator address. 

To get more details about the smart contract, such as the contract's source code and the contract's activity level, we will use Etherscan, an Ethereum explorer. 
The data on Etherscan is updated in real-time, as it syncs with nodes configured on the Ethereum network. Based on each contract's address extracted from Ethereum ETL, we will query Etherscan for the contract address and extract the contract's source code and activity level, if available. There are unverified smart contracts in Etherscan, which means that the developer didn't provide the source code of the smart contract. In this study, we will not consider these smart contracts.

\subsection{Data Collection and Preprocessing}
In this study, to answer the research questions, we will preprocess the collected smart contracts data as follows:
\begin{itemize}
\item Remove comments and white spaces: This will help in reducing the size of collected data and normalizing these data by removing any inconsistencies in structure and format.
 \item Group smart contract duplicates: According to the studies in ~\cite{pinna2019massive, pierro2020organized, oliva2020exploratory}, there are many duplicates in the deployed smart contracts. In this study, we will identify exact duplicates, group them, and consider only one representative to facilitate the analysis process of the source codes. This step is done after removing white spaces and comments.
 \item Reformat multi-file smart contracts: Smart contracts can be coded using one or multiple files. Since we will compare different versions of the smart contracts, it is essential to have each smart contract in a single file. For this purpose, we merge files and remove imports within files for multi-file smart contracts.
\end{itemize}

\subsection{Identifying Upgradeable Smart Contracts}
To answer RQ1, we need to identify upgradeable smart contracts from the collected data. Different patterns exist to upgrade contracts, as mentioned in Section \ref{SMU}. 
We will focus on upgrading patterns introduced by the Ethereum community, \footnote{\url{https://ethereum.org/en/developers/docs/smart-contracts/upgrading/}} and OpenZeppelin.\footnote{\url{https://blog.openzeppelin.com/the-state-of-smart-contract-upgrades/}}. The approaches include data separation, proxy patterns (such as diamond, and UUPS), and strategy pattern.

To detect upgradeable contracts for each existing upgrading approach, we identify a set of policies that distinguish it from the other techniques. These policies include a combination of regular expressions specific to the upgrading pattern as well as the code structure of the pattern. For example, the proxy pattern is identified with the regular expression \textit{delegatecall}. However, not all proxy contracts are upgradeable contracts; forward proxy contracts are used to direct requests to other contracts and not to upgrade a smart contract. To distinguish these two types, it is necessary to analyze the code structure. Upgradeable proxy contracts usually include methods to upgrade the contract address. In contrast, the forward proxy does not have this method, and the requests are delegated to a fixed contract address, as shown in Listings \ref{lst1} and \ref{lst2}.
Furthermore, different upgradeable proxy types exist, such as Diamond proxy and UPPS. The code structure of these types, including the storage structure, differs, which enables us to identify them.

After identifying upgradeable contracts, we statistically analyze how prevalent upgradeable contracts are and what is the most commonly used upgrading pattern. By the end of this step, we will have a list of upgradeable contracts, their address, and upgrading patterns.

\begin{lstlisting}[caption=Upgradeable proxy contract example,
  label=lst1]
pragma solidity ^0.8.0;

contract Proxy {
    address public implementation;

    constructor() public {
        implementation = address(new Implementation());
    }

    |\tikzmark{startb}|function upgradeTo(address _implementation) public|\tikzmark{endb}| {
        require(msg.sender == msg.sender, "Only the owner can upgrade the contract");
        implementation = _implementation;
    }

    function execute(bytes memory _data) public {
        require(implementation != address(0), "Implementation contract not set");
        (bool success, bytes memory returnData) = address(implementation).|\tikzmark{starta}|delegatecall(_dat|\tikzmark{enda}|a);
        require(success, "Execution failed");
    }
}
\end{lstlisting}
\begin{tikzpicture}[remember picture,overlay]
\draw[red,rounded corners]
  ([shift={(-3pt,2ex)}]pic cs:starta) 
    rectangle 
  ([shift={(3pt,-0.65ex)}]pic cs:enda);
\draw[red,rounded corners]
  ([shift={(-3pt,2ex)}]pic cs:startb) 
    rectangle 
  ([shift={(3pt,-0.65ex)}]pic cs:endb);
  \draw[red,rounded corners]
  ([shift={(-3pt,2ex)}]pic cs:startc) 
    rectangle 
  ([shift={(3pt,-0.65ex)}]pic cs:endc);
\end{tikzpicture}

\begin{lstlisting}[caption= Forward proxy contract example,
  label=lst2]
pragma solidity ^0.8.0;

contract Proxy {
    address public implementation;

    constructor(address _implementation) public {
        implementation = _implementation;
    }

    function execute(bytes memory _data) public {
        require(implementation != address(0), "Implementation contract not set");
        (bool success, bytes memory returnData) = address(implementation).|\tikzmark{startc}|delegatecall(_dat|\tikzmark{endc}|a);
        require(success, "Execution failed");
    }
}

\end{lstlisting}

\subsection{Historical Versions of Smart Contracts}
\begin{figure}[t]
    \centering
    \includegraphics[width=\columnwidth, scale =0.55]{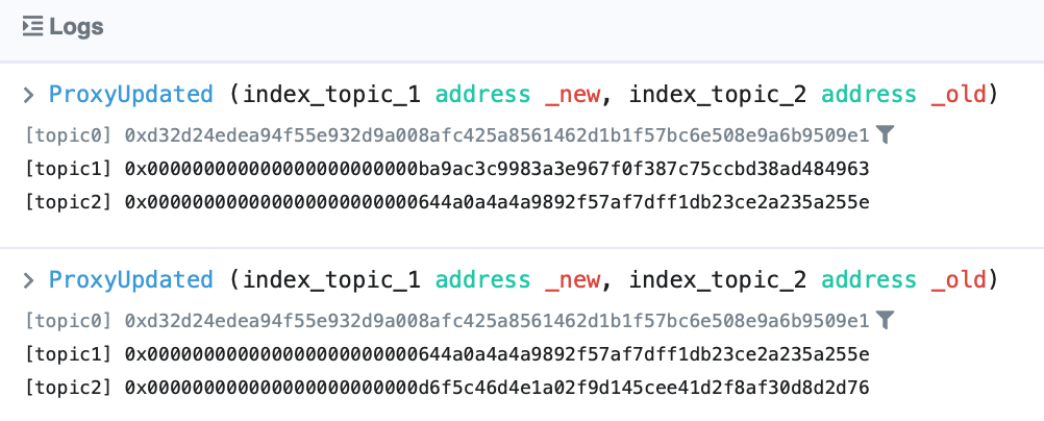}
    \caption{Sample of emitted upgrade contract events}
    \label{fig:event}
\end{figure}

There is no guarantee that all upgradeable contracts will be modified and upgraded. The aim of RQ2 is to analyze how likely an upgradeable contract is to be upgraded.   Hence, we will trace historical versions of each upgradeable smart contract to answer this research question based on the identified upgradeable contracts in RQ1. We will trace versions by analyzing the logs and events of the upgradeable contracts. The logs and historical blocks for each upgradeable contract are obtained from the Ethereum ETL dataset. We will identify the upgrade request (if available) from these logs and save the new version address and details.
For instance, some upgradeable contracts emit events with new contract addresses whenever there is a new upgrade for the smart contract. We trace all the emitted events for the smart contract in these cases and extract the previous version addresses. Figure \ref{fig:event}, shows a sample of emitted events when a contract is upgraded obtained from Etherscan.

\subsection{Identifying the Root Cause of Upgrade}

The objective of RQ3 is to identify the rationale behind upgrading a contract. The upgrade could introduce new features (functions/variables) in the contract, fix code vulnerabilities, or optimize the smart contract in terms of gas and memory. In RQ3, we will consider three main causes of upgrading a contract: fixing vulnerabilities, introducing new features, or optimizing gas cost. If it does not fit these cases, we will consider the root cause as "other" (unknown cause).

To identify the root cause of an upgrade, it's necessary to find the code changes in each smart contract version (upgrade). 
For code changes detection between the identified smart contract versions (from RQ2),  we will use Git dif,\footnote{\url{https://git-scm.com/docs/git-diff}}  a popular and powerful tool thst is well-suited for comparing different code versions.

The Git-diff tool results will help us locate the code changes for each upgrade. We will analyze any security issues in the old and new versions to conclude whether the upgrade was to fix vulnerabilities. This analysis will include using SmartBugs~\cite{ferreira2020smartbugs} to detect vulnerabilities in smart contract versions. SmartBugs framework\footnote{\url{https://github.com/smartbugs/smartbugs}}  offers a wide range of security tools (19 supported tools) that cover different aspects of smart contract security, from static and dynamic analysis to formal verification and optimization. Using the combination of these tools will enable us to thoroughly analysis of smart contracts security.

Furthermore, we analyze if there was gas optimization in the new versions of the contract. In this step, we estimate and compare the gas cost of the contract versions. The gas cost includes the gas fees for contract deployment and contract methods.
\subsection{Upgradeability Impact on Contract's Activity level}
To analyze the impact of upgrading the contract on its usage (RQ4), we will use the received transaction numbers for each version to get insight into the activity level. The transaction number is exported from collected data from Etherscan. Since smart contract versions can have different lifetimes, it is necessary to analyze the contract's activity level while considering its lifetime. The contract's lifetime is measured from when the contract was deployed until a new version was created. If there is no upgraded version, we consider the time we collected the data from Etherscan.
To answer RQ4, we will use a regression model with transaction numbers as the dependent variable and version lifetime as an independent variable. By including version lifetime in the model, we will control for any differences in activity level due to the different lifetimes of smart contract versions. 

\section{Execution Plan and Preliminary Results}
\label{EPPR}
This section presents the execution plan for the study and preliminary results for different stages of the study methods, including data collection, identifying upgradeable contracts, and tracing smart contract versions. 
\subsection{Data Collection}
We have started to collect smart contracts, and so far we have collected around 30M deployed smart contracts up to December 2022. There are many duplicated smart contracts, however, as there are only about 400K unique contracts for the 30M collected contracts. For each smart contract, we have collected the following metadata: the smart contract address, the contract creator address, the timestamp, the compiler version, and the solidity version. Moreover, we have collected from Etherscan the contract source code, if available, and the number of received transactions.

\subsection{Identify Upgradeable Smart Contracts}
\label{IUSC}
\begin{figure}[t]
    \centering
    \includegraphics[scale=0.55]{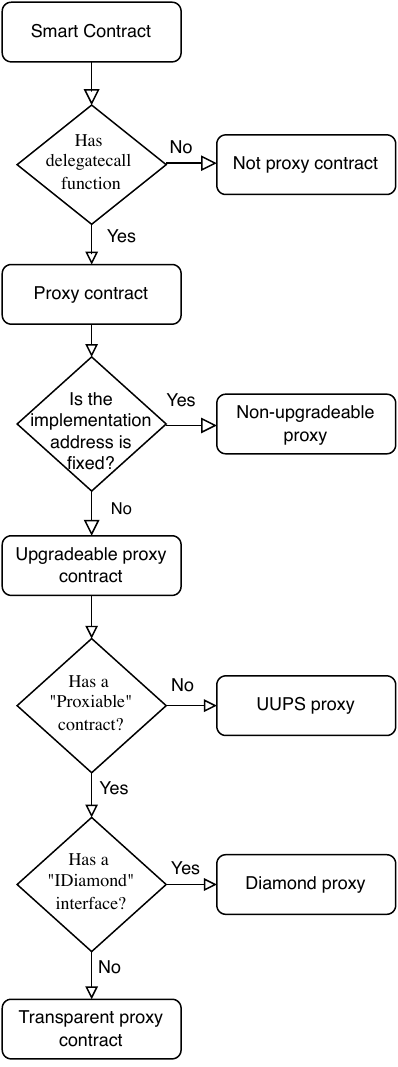}
    \caption{Flow chart to distinguish different  upgradeable proxy contracts}
    \label{fig:proxy}
\end{figure}

This step addresses RQ1 based on the collected data and includes the following steps:
\begin{enumerate}
    \item We design policies to identify and distinguish different approaches to upgrading smart contracts. So far, we have focused on the proxy pattern and developed guidelines to specify proxy contracts. Figure \ref{fig:proxy} illustrates a flow chart to determine different upgradeable proxy contracts. We will extend this flowchart to include other upgrading approaches. 
    \item We identify upgradeable smart contracts that apply specific approaches to upgrade the contract. Currently, we have identified 27K proxy contracts, where 51\% are upgradeable proxy contracts. 
    \item Based on the results of step two, we answer RaQ1, which indicates how prevalent are upgradeable smart contracts. The results will include a list of upgradeable contract addresses and which upgrading approach was applied.
\end{enumerate}
\subsection{Historical Versions of Smart Contracts}
\label{SCHV}
In this step, we further analyze the upgradeable smart contracts identified in \ref{IUSC}, to check whether these contracts were upgraded (new version was created) and trace the upgraded versions. This includes the following steps:
\begin{enumerate}
    \item For each upgradeable smart contract, we get the logs and historical transactions that are initiated by the contract.
    \item We identify upgrading requesst and events for the smart contracts and store the upgrade details based on the logs. Not all upgrade requests are the same, and they can be dependent on the upgrading approaches used. Different upgrading patterns may use different types of upgrade requests. It is important to analyze the upgrade logs and compare them to the different upgrading patterns used to identify the different types of upgrade requests. For example, some upgrading approaches (such as Diamond and UUPS proxy) may require a specific request format (in terms of events and called methods) or include particular keywords, which can help distinguish them from other upgrade requests. On the other hand, other approaches might follow the structure of a similar request depending on the used library (such as OppenZeppelin upgrading libraries \footnote{\url{https://docs.openzeppelin.com/contracts/4.x/api/proxy}}). Hence we first classify the contracts that follow certain standards or use a specific upgrade library.
    
     If the contract does not follow any standard, we identify the contract versions based on the upgrading approach. For instance, in the case of proxy contracts, we can filter the delegatecall requests from the logs and extract the contract addresses. For other approaches, like the data separation approach, we identify the different types of requests when identifying the policies for the approach. These approaches don't use unique functions to upgrade the contract(such as delegetecall), which can be filtered from the contract logs. Hence, the developer must identify an array of version addresses or emit an event whenever an upgrade occurs to trace smart contract versions. 
    
\end{enumerate}

We have randomly sampled 100 upgradeable contracts and traced their upgrades and versions. From 100 contracts, 32 contracts were upgraded, resulting in 76 overall versions. A list of smart contract versions for these 100 contracts is available publicly.\footnote{\url{https://zenodo.org/record/7734661}}

\subsection{Identifying the Root Cause of Upgrade}
This step addresses RQ3, which aims to identify the rationale behind upgrading smart contracts. In this step, we will use the smart contract versions from RQ2 to analyze the root cause of their upgrade. We will consider three main causes of upgrading a contract: fixing vulnerabilities, introducing new features, or optimizing gas cost. If it does not fit these cases, we will label the change as "other" (unknown cause). We might have one or more root causes for upgrading the contract.
We will follow these steps to identify the rationale behind the upgrade:
\begin{enumerate}
    \item We first locate changes between smart contract versions. To identify the changes, we will use the Git diff tool. 
    \item Run the smartBugs tool on both versions of the smart contract to detect any security vulnerabilities. If there is a security vulnerability in the first version that was not found in the second version, mark it as a bug fix.
    \item Identify any changes that were made between the two versions that do not address a security vulnerability. If the second version added lines of code that do not fix a bug, mark it as a new feature. Else if there were functions or variables removed without fixing any vulnerabilities in the code label it as "other" (unknown root cause).
    \item Compare the gas costs of both versions of the smart contract. If the gas cost has decreased in the second version, mark it as a gas optimization.
\end{enumerate}

Algorithm~\ref{alg:root_causes} demonstrates the labeling process of root cause based on the above steps.

\begin{algorithm}
\caption{Identify Root Causes of Smart Contract Upgrade}
\label{alg:root_causes}

\begin{algorithmic}[1]
\REQUIRE $version1$, $version2$
\ENSURE $root\_causes$
\STATE $\mbox{\em versions\_diff} \leftarrow \mbox{\em git\_diff}(version1, version2)$
\STATE $vulnerabilities1 \leftarrow smartbugs(version1)$
\STATE $vulnerabilities2 \leftarrow smartbugs(version2)$
\STATE $\mbox{\em bug\_fixes} \leftarrow []$
\STATE $\mbox{\em new\_features} \leftarrow []$
\STATE $other \leftarrow []$
\STATE $gas\_optimizations \leftarrow []$
\FOR{vulnerability in vulnerabilities1}
\IF{vulnerability $\notin$ vulnerabilities2}
\STATE $\mbox{\em bug\_fixes}.append(vulnerability)$
\ENDIF
\ENDFOR
\FOR{line in $\mbox{\em versions\_diff}.added_lines$}
\IF{line $\notin$ vulnerabilities1 \textbf{and} line $\notin$ vulnerabilities2}
\STATE $\mbox{\em new\_features}.append(line)$
\ENDIF
\ENDFOR
\FOR{line in $\mbox{\em versions\_diff}.removed_lines$}
\IF{line $\notin$ vulnerabilities1 \textbf{and} line $\notin$ vulnerabilities2}
\STATE $other.append(line)$
\ENDIF
\ENDFOR
\STATE $gas\_cost1 \leftarrow calculate\_gas\_cost(version1)$
\STATE $gas\_cost2 \leftarrow calculate\_gas\_cost(version2)$
\IF{$gas\_cost2 < gas\_cost1$}
\STATE $gas\_optimizations.append("\mbox{\em Gas cost decreased}")$
\ENDIF
\STATE $root\_causes \leftarrow []$
\IF{len($\mbox{\em bug\_fixes}$) $>$ 0}
\STATE $root\_causes.append("\mbox{\em Bug fix}")$
\ENDIF
\IF{len($\mbox{\em new\_features}$) $>$ 0}
\STATE $root\_causes.append("\mbox{\em New feature}")$
\ENDIF
\IF{len($gas\_optimizations$) $>$ 0}
\STATE $root\_causes.append("\mbox{\em Gas optimization}")$
\ENDIF
\IF{len($other$) $>$ 0}
\STATE $root\_causes.append("Other")$
\ENDIF
\RETURN $root\_causes$
\end{algorithmic}
\end{algorithm}


\subsection{Upgradeability Impact on Contract's Activity level}

In this step, we will investigate the impact of upgrading a smart contract on the user usage of the contract. This step depends on identified smart contract versions in Section \ref{SCHV}, the number of received transactions for each upgrade, and the age of each contract. The age or lifetime is measured using the following equation:
\begin{align}
\label{eq:1}
  \mbox{\em Age}(V_a) = DT(V_{a+1}) - DT(V_a)
\end{align}
where  $V_a$ is the target version fow which we want to calculate the lifetime $\mbox{\em Age}(V_a)$, $V_{a+1}$ is the next version, and $DT(V)$ is the deployment time of version $V$. If $V_a$ is the latest version, we consider the collection date instead of  $DT(V_{a+1})$ as the end of the deployment period.

To calculate the activity level of each contract, we will use a regression model suitable for the data and calculate a regression equation based on the data. 
Based on the calculated activity levels, we will analyze the impact of upgrading on the activity levels of different versions of the contract.

\section{Related Work}
\label{RW}
There are only a handful of studies focused on the upgradeability of smart contracts. This section presents and discusses these existing studies for smart contract upgradeability approaches. 

Salehi et al.~\cite{salehi2022not} analyzed and evaluated smart contract upgradability patterns. The authors presented a framework for measuring the number of Ethereum upgradeable contracts which utilize certain well-known upgrade patterns. Furthermore, they have analyzed how access control is implemented over smart contract upgradeability.

Bui et al.~\cite{bui2021evaluating} proposed a Comprehensive-Data-Proxy pattern to upgrade smart contracts while enhancing security resilience and scalability. The authors investigated three popular Ethereum attacks that affect smart contract upgradeability: cross-function Reentrancy, typical Reentrancy, and DAO attacks. The presented pattern improves resilience against such attacks compared to the previous upgrading approaches.

Chen et al.~\cite{chen2020finding} introduced a deep learning-based method to detect security issues in the updated version of a destructed smart contract. A contract can be destroyed on Ethereum only by using the Selfdestruct function, which transfers all the Ethers on the contract balance to the contract owner. The authors compare the historical version of destructed contracts and investigated whether security issues were detected in the destructed contracts. 

Fröwis et al.~\cite{frowis2022not} investigated the impact and evaluated the CREATE2 instruction adoption in Ethereum smart contracts. The CREATE2 instruction allows the contract to be modified after deployment on a given address. Furthermore, the authors identified several use cases and attack vectors for the CREATE2 instruction.

Most of the discussed research studies focused on a specific approach to upgrade contracts or proposed a new mechanism to upgrade smart contracts. Only Salehi et al.~\cite{salehi2022not} explored some of the existing approaches to upgrade contracts. 
However, there is a lack of research in analyzing and finding evidence of the existing techniques to upgrade contracts. Moreover, no studies empirically investigated how upgrading patterns are applied or analyzed smart contract versions and the impact of upgrading on the adoption of smart contracts.

\section{Research Challenges and Limitations}
\label{TH}
In this study, we plan to mine all verified smart contracts available in Etherscan. However, the findings in this study can not be generalized to all Ethereum smart contracts, as there are non-verified contracts with no source code available publically. Moreover, we target Solidity, a smart programming language that is not representative of all existing smart contract programming languages. Nevertheless, the aim of this study is not to build a theory applicable to all smart contracts but to inform developers and researchers of the available evolution patterns and the applied approaches to upgrade smart contracts. 

This study will identify upgradeable contracts based on policies that distinguish between different upgrading approaches. These policies are defined after analyzing and identifying the unique characteristics of each upgrading technique. In some scenarios, there might be a wrong identification for upgrading approach due to researchers' biases and limited knowledge. To mitigate this risk, multiple researchers will independently determine the identification policies from well-documented sources that introduce specific or different upgrading approaches, such as Ethereum development standards, \footnote{\url{https://ethereum.org/en/developers/docs/standards/}} and OpenZeppelin. \footnote{\url{https://www.openzeppelin.com}}
We then compare the identified policies by each researcher to reach a consensus on which policies to consider. Furthermore, we will also validate these policies on small smart contract sample and check the accuracy of the policies manually.

Another limitation we might face in this study is tracing the smart contract versions. To trace changes in smart contract state variables, the developer must either create the state variable as an array or emit an event whenever the state variable is changed. Many contracts do not follow this practice, which limits us from finding the changes in the smart contract state. 



\bibliographystyle{ACM-Reference-Format}
\bibliography{SCV}

\end{document}